\documentclass[aps,twocolumn,amsmath,amssymb,superscriptaddress]{revtex4-1}

\usepackage{graphicx}
\usepackage{amsmath}
\usepackage{amsfonts}
\usepackage{amssymb}
\usepackage{amsthm}
\usepackage{color}
\usepackage{hyperref}

\newcommand{\ket}[1]{|{#1}\rangle}
\newcommand{\bra}[1]{\langle{#1}|}
\newcommand{\bracket}[2]{\langle{#1}|{#2}\rangle}

\newcommand{\beq}{\begin{equation}}
\newcommand{\eeq}{\end{equation}}

\renewcommand{\H}{\mathcal{H}}

\begin{document}

\author{Christian Arenz} 
\affiliation{Institute of Mathematics, Physics, and Computer Science, Aberystwyth University, Aberystwyth SY23 2BZ, UK}

\author{Robin Hillier} 
\affiliation{Department of Mathematics and Statistics, Lancaster University, Lancaster LA1 4YF, UK}

\author{Martin Fraas} 
\affiliation{Mathematisches Institut der Universit\"{a}t M\"{u}nchen, Theresianstrasse 39, D-80333 M\"{u}nchen, Germany}

\author{Daniel Burgarth} 
\affiliation{Institute of Mathematics, Physics, and Computer Science, Aberystwyth University, Aberystwyth SY23 2BZ, UK}

\title{Distinguishing decoherence from alternative quantum theories 
by dynamical decoupling}

\date{\today}

\begin{abstract}
A long standing challenge in the foundations of quantum mechanics is the verification of alternative collapse theories despite their mathematical similarity to decoherence. To this end, we suggest a novel method based on dynamical decoupling.  Experimental observation of non-zero saturation of the decoupling error in the limit of fast decoupling operations can provide evidence for alternative quantum theories. The low  decay rates predicted by collapse models are challenging, but high fidelity measurements as well as recent advances in decoupling schemes for qubits let us explore a similar parameter regime to experiments based on macroscopic superpositions.
As part of the analysis we prove that unbounded Hamiltonians can be perfectly decoupled. We demonstrate this on a novel dilation of a Lindbladian to a fully Hamiltonian model that induces exponential decay.
\end{abstract}

\maketitle

\section{Introduction} 
Despite of its puzzling nature and persistent foundational problems, such as the infamous measurement problem, quantum mechanics remains one of the most precise and successful physical theories to date. This makes it hard to develop alternative theories (for an overview we refer to \cite{Bassi1, AdlerBook, Bassi2}), which are either bound to agree with quantum mechanics on all measurable aspects -- and therefore being indistinguishable from it -- or must disagree with it only at the most subtle level, which means that such theories are hard to falsify experimentally. While in our daily life quantum effects do not appear to play a role, this does not imply that it is an incomplete theory, as the onset of classicality can -- at least up to a certain degree \cite{BookSchlosshauser, ArticleSchlosshauser} -- be explained from \emph{within} quantum theory, using the concept of decoherence.

Decoherence arises from the coupling of a quantum object with other degrees of freedom, which washes out quantum mechanical features. Besides being a major obstacle to quantum computing, decoherence is also an obstacle to the tests of theories alternative to quantum mechanics, since it tends to obscure the -- already minimal -- deviations they predict from the usual Schr\"odinger dynamics. Even worse, since most alternative theories aim to explain the onset of classicality, they predict features identical in their mathematical nature to decoherence \cite{Bassi4}. The main aim of this article is to demonstrate that while these models might be \emph{mathematically} identical, they are \emph{physically} distinguishable, irrespectively of decoherence.  At first, this seems impossible. Especially in quantum information theory, the Church of the Larger Hilbert Space -- the idea that any noisy dynamics or state might equally well be represented by a noiseless one on a dilated space -- is so deeply rooted that such a distinction seems heretic.

A method to distinguish decoherence from alternative quantum theories (AQT) which is obvious but impractical is to derive ab initio predictions of decoherence and compare these with experiments. Unfortunately, the predictive power of decoherence models till date is low, as they contain many free parameters to fit. We therefore aim to develop methods which are independent of the details of the decoherence involved, as well as of the specific AQT considered.

Our work is based on a very simple idea, namely that dynamical decoupling \cite{DCBook} -- a popular method to suppress quantum noise -- only works for systems which are \emph{truly} coupled to environments \cite{USMath}, but not for systems which have intrinsic noise terms, as arriving from axiomatic modifications of Schr\"odinger's equation \cite{Dosi, Milburn, Bassi3, Bassi4}.

This seems to leave us with an amazingly simple strategy to distinguish decoherence from AQT: apply decoupling, and if it works, then the noise was due to standard quantum theory; if it does not work, it can provide evidence for AQT. Is this therefore the most successful ``failed'' experiment ever? 
Of course not: we need to be convinced that the experiment did not work \emph{despite good effort}, in other words, we need to know quantitatively how much the experiment can fail while still being in the realms of standard decoherence; and how much it can succeed despite being in the realms of AQT. This poses an additional problem. It is a common view that dynamical decoupling only works for environments inducing non-exponential decay (sometimes referred to as `non-Markovian', although this term is used ambiguously in the literature). This means that if the observed quantum dynamics shows exponential behaviour, we would not be able to distinguish it from AQT. On the other hand, most AQTs predict exponential decay \cite{Bassi4}. 

The reason for this common view is that exponential decay can only be obtained from an unbounded interaction with the environment \cite{OpenQS2}, for which standard error analysis of dynamical decoupling fails \cite{LVioalRandD}. Perhaps surprisingly, we will prove in section \ref{chap:DynUnb} that in general even unbounded Hamiltonians can be decoupled and hence distinguished from intrinsic decoherence. This general proof is illustrated by an analytically solvable example \ref{sec:ShallowPM}. We can conclude that non-exponential dynamics is in general not the underlying mechanism of dynamical decoupling. This result extends the applicability of decoupling to a vast class of system-environment interactions and has applications in quantum engineering beyond the scope of this paper.

Finally, dynamical decoupling arises in the limit of infinitely fast quantum gates, so in practice it is never perfect. How fast should these operations be so that decoherence and AQT can be distinguished? Below, we provide numerical simulations of two common models and asymptotical bounds (referring to \cite{USMath} for a detailed mathematical analysis) regarding these questions. As we will see below, the convergence speed can depend strongly on the initial bath state, which implies that  \emph{model independent bounds}, e.g., depending only on the observed decay rates of the system,  cannot be provided. Nevertheless, experimental evidence can be provided if a saturation of fidelity is observed under increasingly fast operations. For the parameter range explorable by our scheme, we can do the following rough estimate. The strongest intrinsinc decay rates for qubits predicted by AQT are of the order of $10^{-8}\,s^{-1}$  corresponding to a half-life time of several \emph{years} \cite{Bassi3}. Precision measurements of qubits on the other hand are very well developed meaning that coherence decay of the order of percent can be detected. This means that if one aims to keep a qubit from detectable decay for several days, the first AQT models could be detected or excluded. At present qubit coherence times can be prolonged by dynamical decoupling up to six hours \cite{Zhong}. This is still a few orders of magnitude off the theoretical predictions, which is comparable  to the usual AQT tests in the macroscopic superposition regime.

Our results pave the way to test AQT in low-dimensional systems, including qubits, where AQT predicts very weak effects \cite{Bassi1}, but where dynamical decoupling is very efficient, and where accurate tomography can be performed \cite{TOMO}. This is a different parameter regime compared to tests using macroscopic superpositions \cite{MS1,MS2,MS3,MS4}, where AQT predict stronger effects but dynamical decoupling is challenging (see, however, \cite{Vitali}).
\par
\section{ Dynamical decoupling for bounded Hamiltonians}
\label{sec:dynamicald}
Dynamical decoupling is a highly successful strategy to protect quantum systems from decoherence \cite{DCBook}. Its particular strength is that it is applicable even if the details of the system-environment coupling are unknown. In the context of quantum information the theoretical framework was developed in \cite{LVioal1, LVioalRandD} and the efficiency of different decoupling schemes was studied and improved for several environmental models in \cite{DecouplingSpinB1, DecouplingSpinB2, DecouplingSpinBRand, Uhrig1, Uhrig2, Uhrig3}. Many experiments, such as \cite{NVcentre, classicalnoise, QubitinSolid}, demonstrate the applicability of dynamical decoupling in an impressive way by prolonging coherence times a few orders of magnitude. Additionally, dynamical decoupling can be combined with the implementation of quantum gates which makes it a viable option to error correction \cite{DCandEC, DCQgates}. The idea of dynamical decoupling 
is to rapidly rotate the quantum system by means of classical fields to average the system-environment coupling to zero.

More precisely consider the unitary decoupling operations $v$ taken from the set $V$ of $|V|$  unitary $d\times d$ matrices satisfying $\frac{1}{|V|}\sum\limits_{v\in V}vxv^{\dagger}=\frac{1}{d}\text{tr}(x)\openone$ for any matrix $x$. An example of such a set for a single qubit are the Pauli matrices $V=\{\openone,\sigma_x,\sigma_y,\sigma_z\}$. While usually 
dynamical decoupling is discussed in the realm of a unitary time evolution, we already allow a noisy dynamics generated by a Lindbladian $\mathcal{L}$ because we later want to see what happens for AQT. This dynamics is now modified by decoupling operations $v_{i}\in V$ with $i=1,...,n$ applied instantaneously in time steps $\Delta t$. After time $t=n\Delta t$ the system has evolved according to 
\begin{align}
\label{eq:timeevolutiondecoupling}
\Lambda_{t,n}(\cdot)=\prod\limits_{i=1}^{n}\text{Ad}(v_{i})\exp(\Delta t\mathcal{L})\text{Ad}(v_{i}^{\dagger})(\cdot),
\end{align}
where $\text{Ad}(v_{i})(\cdot)=v_{i}(\cdot)v_{i}^{\dagger}$ and the product is time-ordered. The generalization to time-dependent generators is straight forward and will be used later in the examples. Throughout this paper we consider perfect decoupling operations, while bounds for the non-perfect case can be found for example in \cite{Acc1, Acc2, Acc3}. The decoupling operations are chosen uniformly random from $V$, which has some advantage over deterministic schemes \cite{LVioalRandD, DecouplingSpinBRand}. Notice that our definition of random dynamical decoupling differs slightly from \cite{LVioalRandD}. The time evolution \eqref{eq:timeevolutiondecoupling} becomes a stochastic process with expected dynamics determined by
\begin{align}
\label{eq:decouplingcondition}
\bar{\mathcal{L}}:=\frac{1}{|V|}\sum\limits_{v\in V}\text{Ad}(v)\mathcal{L}\text{Ad}(v^{\dagger}).
\end{align}
This leads to the \emph {decoupling condition} $\bar{\mathcal{L}}=0$, which one requires in order to successfully suppress decoherence. Note that this condition is independent of whether we use a deterministic or random decoupling scheme \cite{LVioal1}. The idea behind this condition is that it ensures the cancellation of $\mathcal{L}$ in first order in $\Delta t||\mathcal{L}||$. For $\Delta t \to 0$, keeping the total time $t$ fixed, the time evolution \eqref{eq:timeevolutiondecoupling}  becomes therefore effectively the identity.

 Hamiltonian dynamics $\mathcal{L}(\cdot)=i[H,\cdot]$  can always be supressed through dynamical decoupling. In the section \ref{chap:DynUnb} we prove that this is even true for unbounded Hamiltonians. But what happens for AQT? Note first of all that for AQT models that modify the Schr\"odinger equation in a nonlinear way, it was argued in \cite{Bassi4} that under the assumption of the no-signalling principle the resulting dynamics is  described by a  time independent Lindblad operator 
 \begin{align}
 \mathcal L(\cdot)=\sum\limits_{j=1}^{d^{2}-1}\gamma_{j}(2L_{j}(\cdot)L_{j}^{\dagger}-(L_{j}^{\dagger}L_{j}(\cdot)+(\cdot)L_{j}^{\dagger}L_{j})),
 \end{align}
yielding the averaged Lindbladian 
\begin{align}
\label{eq:averagedL}
\bar{\mathcal L}(\cdot)=\sum\limits_{j=0}^{d^{2}-1}2\gamma_{j}\left(\frac{1}{|V|}\sum\limits_{v\in V}vL_{j}v^{\dagger}(\cdot)vL_{j}^{\dagger}v^{\dagger}-\frac{1}{d}\text{tr}(L_{j}^{\dagger}L_{j})(\cdot)\right).
\end{align} 
We will henceforth refer such  AQT dynamics as \emph{intrinsic decoherence}. In order to avoid confusion, we will write \emph{extrinsic decoherence} for decoherence arising in standard quantum theory. 
Surprisingly if the dynamics includes intrinsic decoherence, the decoupling condition can \emph{never} be fulfilled. Intuitively the irreversible nature of the non-unitary dynamics, i.e. the increase of entropy, makes it impossible to counteract the loss of coherence with unitary decoupling pulses. For a detailed mathematical proof we refer to \cite{USMath}. This is a remarkable result since it enables us to distinguish  two different seemingly equal decoherence mechanisms. We remark that the generalization to time-dependent Lindbladians is straightforward allowing our technique also to discriminate non-exponential collapse models from extrinsic decoherence.

In the limit of arbitrarily fast decoupling operations $(\Delta t\to 0)$ dynamical decoupling works perfectly for extrinsic decoherence.  However, in practice even   dynamical decoupling of extrinsic decoherence can never be perfect meaning that higher orders in $\Delta t||\mathcal{L}||$ enter the resulting dynamics. To detect the presence of intrinsic decoherence we therefore need to develop an extrapolation for $ \Delta t\to 0$. Furthermore to distinguish extrinsic and intrinsic decoherence we need bounds. Using a central limit theorem, such bounds are developed in \cite{USMath} for the expectation of the decoupling error $\bar{\epsilon}$, while here we will focus on specific examples. The decoupling error $\epsilon=\text{tr}\{(\openone-\Lambda_{t,n})^{\dagger}(\openone-\Lambda_{t,n})\}/d^2$ compares the free evolution under random dynamical decoupling with the identity operation. In the limit $\Delta t\to 0$, keeping the total time $t$ fixed, the decoupling error becomes \cite{USMath},
\begin{align}
\label{eq:averagedDecError}
\epsilon=\frac{1}{d^{2}}\text{tr}\left( (\openone-\exp(\bar{\mathcal L}t))^{\dagger} (\openone-\exp(\bar{\mathcal L}t))    \right),
\end{align}
where for extrinsic decoherence the time evolution of the total system is followed by the partial trace over the environment yielding $\epsilon=0$ for $\Delta t\to 0$. Note that the decoupling error can be estimated in an experiment by performing process tomography \cite{NielsenProcessT}. Simpler ¨fingerprints¨ to distinguish AQT which do not require process tomography can easily be derived for specific systems. In the following we emphasize the physics calculating bounds for two common models.  

\section{Models and bounds} 
To demonstrate our method we consider two different types of decoherence of a single qubit, namely amplitude damping and pure dephasing.
\subsection{Two qubit model}
To begin with suppose that one observes a dynamics described by an amplitude damping (AD) channel, given by the Lindblad operator 
\begin{align}
\label{eq:LindbladienAmpD}
\mathcal{L}_{\text{AD}}(\cdot)=-\gamma(\sigma_{+}\sigma_{-}(\cdot)
+(\cdot)\sigma_{+}\sigma_{-}
-2\sigma_{-} (\cdot)\sigma_{+}),
\end{align}
with $\sigma_{\pm}$ the raising and lowering Pauli operators. Within the extrinsic decoherence model such amplitude damping dynamics can be obtained by a time dependent interaction with an ancilla qubit $(A)$ initialized in its ground state. The total Hamiltonian reads 
\begin{align}
\label{eq:twoqubitH}
H(t)=g(t)(\sigma_{+}\otimes\sigma^{(A)}_{-}+\sigma_{-}\otimes \sigma^{(A)}_{+}),
\end{align}
 with the time dependent coupling constant  $g(t)=\gamma/\sqrt{\exp(2\gamma t)-1}$. The Hamiltonian $H(t)$ commutes with itself at all times such that the time evolution of the composite system can easily be integrated. After tracing over the ancilla qubit one obtains precisely the two Kraus operators which describe the amplitude damping channel generated by \eqref{eq:LindbladienAmpD}. Note that at $t=0$ the interaction strength $g(t)$ diverges while the time evolution operator remains well defined. Clearly there are other possible choices of the system-bath Hamiltonian that lead to the same dynamics. For example within the Born-Markov approximation the same Lindblad operator \eqref{eq:LindbladienAmpD} is obtained by a time independent interaction of the qubit with a bath of harmonic oscillators at zero temperature. However as a toy model, $\eqref{eq:twoqubitH}$ has the advantage of being simpler. Such time-dependent dilations may also find applications in other context.

\begin{figure}[!h] \includegraphics[width=0.8\columnwidth]{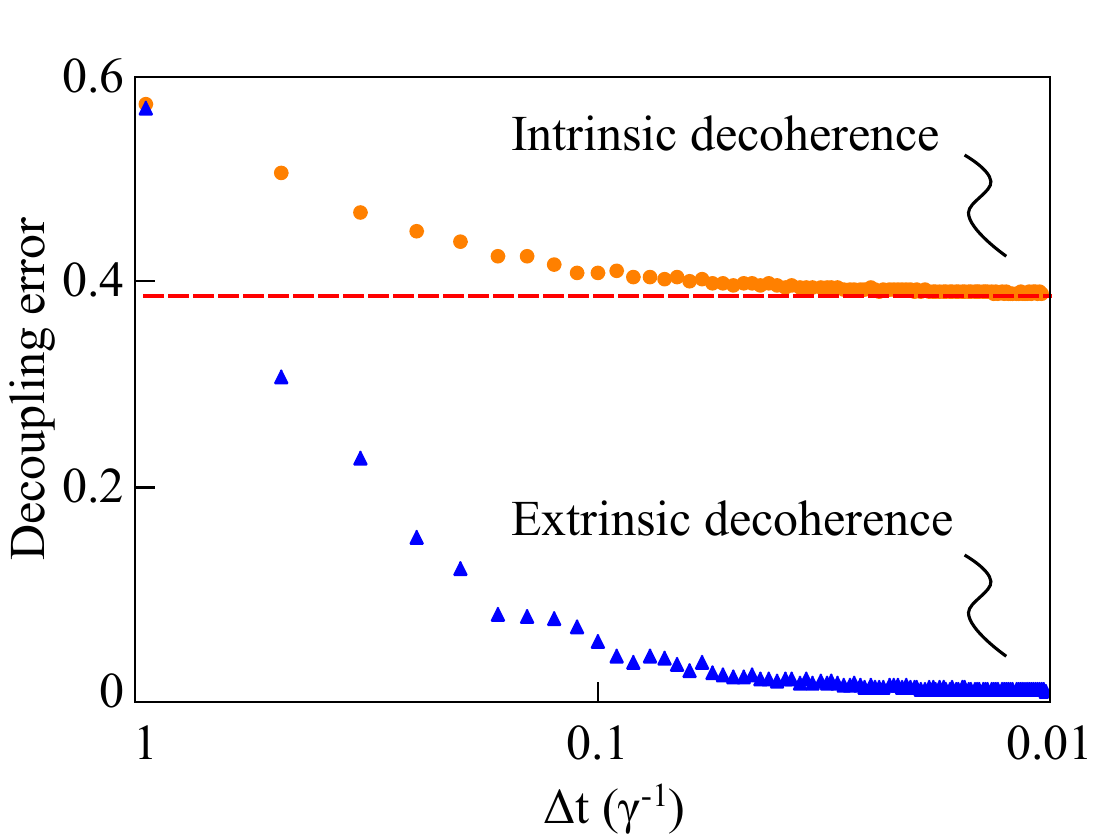}
  \caption{\label{fig:twoquibtimodel} (Colour online) Averaged decoupling error under random dynamical decoupling as a function of $\Delta t$ on an inverse logarithmic scale for the total time  $t=\gamma^{-1}$. The circles correspond to pure intrinsic decoherence described by \eqref{eq:LindbladienAmpD}, the triangles to extrinsic decoherence given by \eqref{eq:twoqubitH} and the dashed line shows the asymptotic behavior \eqref{eq:lowerboundAD} for the intrinsic decoherence case for $\Delta t \to 0$. The average was taken over $100$ trajectories.}
\end{figure}

Now we turn to the question how well dynamical decoupling can distinguish between extrinsic decoherence, given by the Hamiltonian \eqref{eq:twoqubitH}, and pure intrinsic decoherence given by the Lindbladian \eqref{eq:LindbladienAmpD}. Using \eqref{eq:averagedL} one finds for the intrinsic decoherence case the averaged Lindblad operator $\bar{\mathcal{L}}_{\text{AD}}(\cdot)=-\gamma(\openone(\cdot)-\sigma_{-}(\cdot)\sigma_{+}
-\sigma_{+}(\cdot)\sigma_{-})$ which determines the dynamics in the limit of infinitely fast decoupling operations. The first observation is that $\bar{\mathcal{L}}_{\text{AD}}$ does not vanish. With \eqref{eq:averagedDecError} we can furthermore derive the following asymptotic behaviour for the decoupling error in the intrinsic decoherence case 
\begin{align}
\label{eq:lowerboundAD}
\epsilon_{\text{AD}}^{\text{int}}\to \frac{1}{4}\left(3-e^{-\gamma t}\left(4-e^{-3\gamma t}\right) \right), \quad \Delta t\to 0,
\end{align} 
and for $\gamma t\gg 1$ it approaches a value of $3/4$. In Fig. \ref{fig:twoquibtimodel} we evaluated the averaged decoupling error for intrinsic and extrinsic decoherence as a function of $\Delta t$ for a fixed total time $t=\gamma^{-1}$.  
We see that for the Hamiltonian model \eqref{eq:twoqubitH} the decoupling error tends to zero. The asymptotic behaviour of the averaged trajectories allows us to distinguish intrinsic from extrinsic decoherence: for purely intrinsic decoherence we have \eqref{eq:lowerboundAD}, while for purely extrinsic it is $0$, and everything in-between must correspond to a mixture of the two. The actual speed of convergence to the limit in the extrinsic case depends on the chosen dilation \cite{LVioalRandD}, so that we cannot say how small $\Delta t$ has to be chosen in order to distinguish with certainty.
\subsection{Spin-boson model}
Next, we consider a more realistic and experimentally relevant model describing pure dephasing (PD) in the $\sigma_{z}$ basis of the qubit. The Lindbladian reads 
\begin{align}
\label{eq:LindbladienDP}
\mathcal{L}_{\text{PD}}(t)(\cdot)&=-\frac{\gamma(t)}{4}[\sigma_{z},[\sigma_{z},\cdot\,]],
\end{align} 
where the time dependent damping rate $\gamma(t)$ will be specified later. 
As extrinsic decoherence such PD would arise from an interaction with a bosonic heat bath given by
\begin{align}
\label{eq:spinbosonH}
H=\sum\limits_{k}\omega_{k}a_{k}^{\dagger}a_{k}+\sigma_{z}\sum\limits_{k}(g_{k}a_{k}^{\dagger}+g_{k}^{*}a_{k}),
\end{align}
where $a_{k}^{\dagger},a_{k}$ are the bosonic creation and annihilation operators of the $k$th field mode and $g_{k}$ are coupling constants quantifying the interaction strength to each harmonic oscillator. After tracing over the bath degrees of freedom \cite{OpenQS,SpinBosonMasterEq1, SpinBosonMasterEq2} one finds for the time dependent damping rate $\gamma(t)=4\int_{0}^{t}ds\int_{0}^{\infty}d\omega I(\omega)\coth\left(\frac{\omega}{2T}\right)\cos(\omega s)$ where the continuum limit was performed and the spectral density $I(\omega)$, which contains the statistical properties of the bath, and the temperature $T$ of the bath were introduced.

For an intrinsic dephasing mechanism given by \eqref{eq:LindbladienDP} the decoupling operations $V$ do not affect the dynamics $v\sigma_{z}v^{\dagger}=\pm\sigma_{z}$ for all $v\in V$ such that  $\mathcal{L}_{\text{PD}}=\bar{\mathcal{L}}_{\text{PD}}$. Therefore the decoupling error in the intrinsic decoherence case is 
governed by the dynamics generated by $\mathcal{L}_{\text{PD}}$ and with $\epsilon_{\text{PD}}=\frac{1}{4}\text{tr}\left((\openone-\exp(\int_{0}^{t} dt^{\prime}\mathcal L_{\text{PD}}(t^{\prime}) ))^{\dagger}(\openone-\exp(\int_{0}^{t} dt^{\prime}\mathcal L_{\text{PD}}(t^{\prime}) )) \right)$  one finds
independently of $\Delta t$,  
\begin{align}
\label{eq:boundDP}
\epsilon_{\text{PD}}^{\text{int}}=\frac{1}{2}\left[1-\exp\left(-\int_{0}^{t}\gamma(t^{\prime})
dt^{\prime}\right)\right]^{2},
\end{align}
showing that the asymptotic decoupling error is given by $1/2$. Based on the spin-boson Hamiltonian \eqref{eq:spinbosonH}
it was shown in \cite{DecouplingSpinBRand} that under random dynamical decoupling the spectral density gets renormalized by a factor that ensures for $\Delta t\to 0$ the suppression of decoherence.

\begin{figure}[!h] \includegraphics[width=0.8\columnwidth]{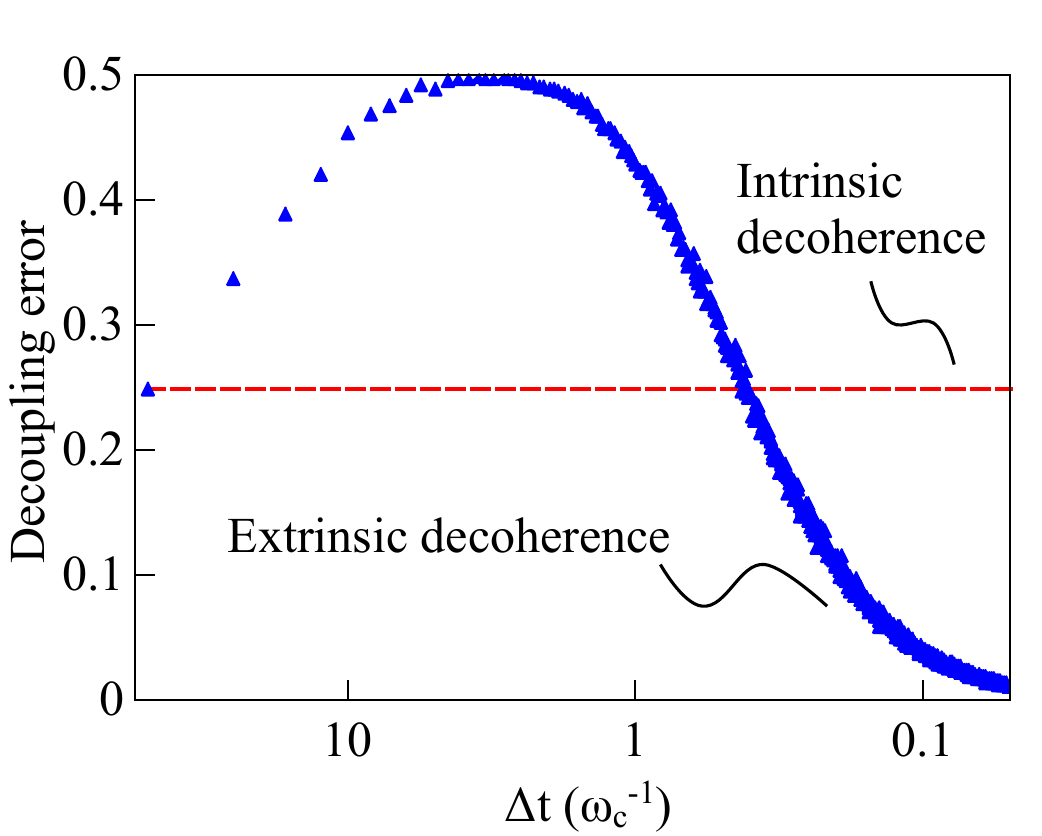}
  \caption{\label{fig:spinbosonlowtemp} (Colour online) Averaged decoupling error under random dynamical decoupling as a function of $\Delta t$ on an inverse logarithmic scale  evaluated for $t=50\,\omega_{c}^{-1}$.  The triangles correspond to extrinsic decoherence given by the spin boson model \eqref{eq:spinbosonH} where the dashed line corresponds to intrinsic decoherence \eqref{eq:LindbladienDP} which is independent of $\Delta t$ here \eqref{eq:boundDP}. The average was taken over $100$ trajectories.}
\end{figure}
 
Because the decoupling operations $V$ give the same spectral density as in \cite{DecouplingSpinBRand} we can easily evaluate the averaged decoupling error for extrinsic and intrinsic decoherence ( Fig. \ref{fig:spinbosonlowtemp} ).  We chose an ohmic spectral density with a sharp cut off $I(\omega)=1/4\kappa\omega\theta(\omega-\omega_{c})$ with $\kappa=0.25$ a measure of the coupling strength to the environment and $\omega_{c}=100$ the cut off frequency. We calculated the averaged decoupling error in the low temperature limit $\omega_{c}/T=10^{2}$. 

Note that for $\Delta t \gtrsim 0.5\,\omega_{c}^{-1}$ decoherence gets accelerated as reported in \cite{DecouplingSpinBRand} in the extrinsic case  since  the decoupling error is higher than the decoupling error  that is obtained for the dynamics generated by $\mathcal{L}_{PD}$. 

\section{Dynamical decoupling of unbounded Hamiltonians}
\label{chap:DynUnb}

 Many physical environments are modelled as infinite dimensional system, often with \emph{unbounded} interactions. In order to discuss dynamical decoupling of such systems, we find it enlightening to start with a specific, analytically solvable model, before providing a general proof that generally even unbounded time-independent Hamiltonians can be decoupled.

\subsection{Shallow pocket model}
\label{sec:ShallowPM}
We now provide an analytically solvable model of an unbounded, time-independent Hamiltonian which, without approximations, leads to a time-independent dephasing Lindbladian, but can be decoupled arbitrarily well. It is an example of an exact time-independent dilation describing a small system coupled to a fictitious particle on a line.  After tracing over the decrees of freedom of the particle we obtain a time independent Lindblad generator for the reduced dynamics of the system. The particle cannot store energy internally -- hence the name -- and the dynamics is governed by an interaction Hamiltonian
\begin{align}
H  = \frac{g}{2}\sigma_z\otimes x = \frac{g}{2} \left( \begin{array}{cc} x & 0 \\ 0& -x \end{array} \right),
\end{align}
where $x$ is the position operator and the small system is a qubit for simplicity and $g$ a coupling constant. The Hamiltonian is diagonal and the evolution of a joint density matrix is
\begin{align}
\rho(t,x) = \left( \begin{array}{cc} \rho_{11}(0,x) & \rho_{10}(0,x) e^{i g x t} \\
						        c.c. & \rho_{00}(0,x) \end{array} \right).
\end{align}
A reduced dynamic displaying exponential decay is achieved by choosing an initial state $\rho \otimes \ket{\psi} \bra{\psi}$ where
\begin{align}
\bracket{x}{\psi} =  \sqrt{\frac{\gamma}{\pi}} \frac{1}{x + i \gamma}.
\end{align}
After integrating out the particle degree of freedom we obtain, through the Fourier transform of a Lorentzian, a purely exponential decay of the off diagonal terms,
\begin{align}
\rho(t) = \left( \begin{array}{cc} \rho_{11}(0) & \rho_{10}(0) e^{-  g \gamma t} \\
						        c.c. & \rho_{00}(0) \end{array} \right),
\end{align}
 which corresponds to a time-independent dephasing Lindbladian 
\begin{align} 
\label{eq:dpL} 
 \mathcal{L}(\cdot)=-g\frac{\gamma}{4}[\sigma_z,[\sigma_z,\cdot]].
\end{align}
The model can be perfectly decoupled using $\mathbb{Z}_2$ controls $v_0 = \openone,\,v_1 = \sigma_x$. In fact $v_1 H v_1^\dagger = - H$ and hence
\begin{align}
\label{eq:shallowpdecoupling}
v_0 \exp(i \Delta t H) v_0^\dagger v_1 \exp(i \Delta t H) v_1^\dagger = \openone.
\end{align}
\begin{figure}[!h] \includegraphics[width=0.8\columnwidth]{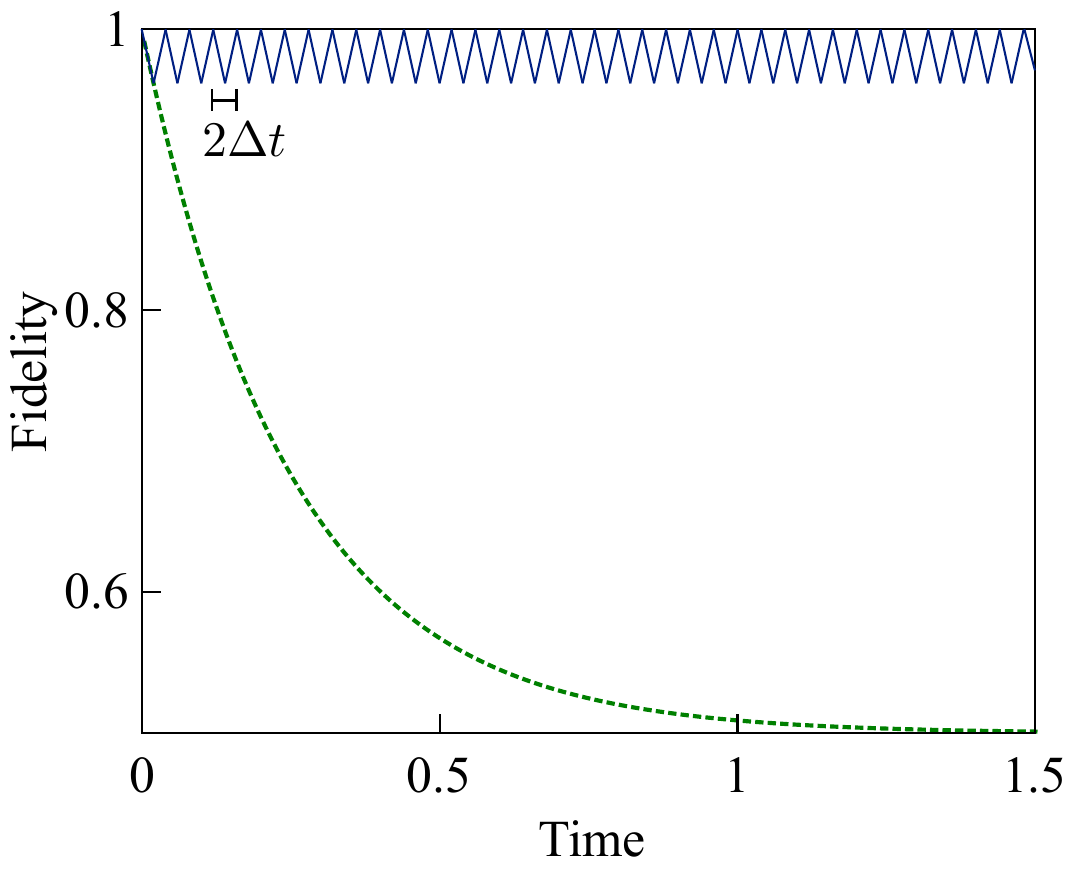}
  \caption{\label{fig:shallowpocketmodel} (Colour online) Schematic representation of the fidelity for exponential dephasing (dotted green line) to stay in a coherent superposition of ground and excited state. The solid blue line shows the dynamics of the qubit under dynamical decoupling.}
\end{figure}
 This model displays similar effects as the above ones, which means that the explicit time-dependence of the Hamiltonian/Lindbladian of the first two examples is not relevant to the discussion.
In Fig. \ref{fig:shallowpocketmodel} we show the fidelity $\mathcal F(t)=\frac{1}{2}\left(e^{-g\gamma t}+1\right)$ (dotted green line) of being in a coherent superposition of ground and excited state obtained from the dynamics generated by the Lindbladian \eqref{eq:dpL}. The solid blue line shows the reduced dynamics of the shallow pocket model under dynamical decoupling \eqref{eq:shallowpdecoupling}.  

The shallow pocket model is a counterexample to dynamical decoupling working for non-exponential decay only. For a fixed decoupling time $\tau$ the fidelity never drops below $\mathcal{F}(\tau)$. The model also highlights some of the unpleasant mathematical properties required for modelling strict exponential decay: the initial state of the system is not in the domain of the interaction \cite{Comment on domain initial state}, which in turn is unbounded below and above \cite{OpenQS2}. Such properties indicate that the general proof below requires a certain degree of mathematical precision.

\subsection{General proof}

It is a fact of nature and an ubiquitous challenge in the mathematical treatment of quantum mechanics that unbounded Hamiltonians cannot be defined everywhere \cite[Chapter VIII]{RS}. A definition domain $D(H)$ has to be specified in order to make a clear sense of an unbounded Hamiltonian $H$.  For example the notion of self-adjointness, properties of a sum $H_1 +H_2$, etc has to take the definition domain into account. Starting with a pioneering work of von Neumann a machinery has been developed with a purpose to circumvent these problems when dealing with a derived quantum mechanical phenomena.  This is precisely our case, we show that whenever a Hamiltonian which couples a finite-dimensional system of size $d$  to an infinite-dimensional bath can be reasonably defined then it can be decoupled perfectly.

All Hamiltonians under our consideration have a sum-like structure consisting of the system{\textbackslash}bath free Hamiltonians and the interactions. A core of an operator \cite{RS} is then a natural notion to make sense of this sum in the most general setting. We postpone this technical discussion by few paragraphs and start with a natural -- albeit less general -- setting where this notion is not needed. It includes for example the case when the interaction Hamiltonian is relatively bounded with respect to the free Hamiltonian.

 We assume that a Hamiltonian describing the system is a densely defined self-adjoint operator of the form
 $H = H_S \otimes \openone + \openone  \otimes H_B+ \sum_\alpha S_\alpha \otimes R_\alpha$ on the tensor product Hilbert space $\H_{SB}=\H_S\otimes\H_B$, with $H_B$ itself self-adjoint on a dense domain $D(H_B)$ and $D(H)= \mathbb{C}^d \otimes D(H_B)$.
For simplicity we only consider deterministic decoupling schemes here, while the random case can be proved using \cite[Th.2.2]{Kurtz} (c.f. forthcoming work for details). The announced perfect decoupling of such a Hamilotnian might be surprising given that the usual derivation of dynamical decoupling hinges on a perturbative expansion $\exp(i \Delta t A) \sim \openone + i \Delta t A +O(\Delta t^2)$ and a limit formula
\begin{align}
\left(1 + \frac{A}{n} + O(n^{-2})\right)^n \to \exp(A).
\end{align}
In particular all standard error bounds \cite{LVioalRandD} become infinite for unbounded Hamiltonians. These apparent problems can be circumvented by means of a deep generalization of the above limit formula due to Chernoff \cite{Chernoff}, c.f. also \cite[Chapter 8.]{Nelson}: \emph{Let $F(t), ||F(t)|| \leq 1$ be a family of operators on a Hilbert space $\H$ with $F(0) = \openone$ and suppose that $(F(t) - \openone)(\psi)/t \to A \psi$ as $t\rightarrow 0$, for every $\psi\in\H$ in a core of $A$. Then we have
\begin{align}
\label{eq:Chernoff}
\lim_{n \to \infty} F\left(\frac{t}{n}\right)^n (\psi) = \exp(t A)\psi, \quad \psi\in\H.
\end{align}
}
We apply Chernoffs theorem with $F(t) = \Pi_{v\in V} v \exp(i H t/|V|) v^{\dagger}$ and $H$ as above. Then for $\psi\in D(H)$,
\begin{align}
\label{eq:converCher}
\frac{(F(t) - \openone)(\psi)} {t} \rightarrow i\left(\frac{1}{|V|}\sum_{v\in V} v H v^\dagger\right)\psi= i (\openone\otimes H_B)\psi,
\end{align}
 due to the decoupling property of $V$, as $t\rightarrow 0$ and for every $\psi$ in the domain of all $v H v^\dagger$'s. Note that the convergence in \eqref{eq:converCher} is not obvious since the use of the Taylor series is not well defined for unbounded operators. Along the lines of \cite{Suzuki} it can be proven instead on the group level, by rearranging the exponentials in such a way that Stone's theorem can be used. Consider for example as a system a qubit with $V$ the Pauli group. We can evaluate the limit \eqref{eq:converCher} using
\begin{widetext}
\begin{align}
\frac{(F(t)-\openone)(\psi)}{t}&=\frac{1}{t}\left(e^{-i\sigma_{z}H\sigma_{z}t}-\openone\right)\psi+\frac{1}{t}e^{-i\sigma_{z}H\sigma_{z}t}\left(e^{-i\sigma_{y}H\sigma_{y}t}-\openone\right)\psi+\frac{1}{t}e^{-i\sigma_{z}H\sigma_{z}t}e^{-i\sigma_{y}H\sigma_{y}t}\left(e^{-i\sigma_{x}H\sigma_{x}t}-\openone\right)\psi,\nonumber \\
\label{eq:exampleConverg}
&+\frac{1}{t}e^{-i\sigma_{z}H\sigma_{z}t}e^{-i\sigma_{y}H\sigma_{y}t}e^{-i\sigma_{x}H\sigma_{x}t}\left(e^{-iHt}-\openone\right)\psi,
\end{align} 
\end{widetext}
with $\psi\in \mathbb C^{2}\otimes D(H_{B})$. By assumption all $vHv^{\dagger}$ are self-adjoint on this domain, so we can apply Stone's theorem for each summand of \eqref{eq:exampleConverg} yielding the desired result \eqref{eq:converCher} as $t$ goes to zero.  We conclude that perfect dynamical decoupling
\begin{align}
\lim_{n\rightarrow\infty} \mathrm{tr}_B (\Lambda_{t,n}\rho ))  =  \mathrm{tr}_B \big( e^{i t\openone \otimes H_B} \rho e^{-i t\openone \otimes H_B} \big) 
=  \mathrm{tr}_B (\rho),
\end{align}
is possible where $\rho$ is the density operator of the system and the bath. 

Notice that many examples including the shallow pocket model verify the above assumptions of self-adjointness. Nevertheless, we aim for even bigger generality and to achieve this we introduce the notion of a core into our discussion. A core of an operator is a subspace of its domain such that restriction of the operator to the core and subsequent closure gives back the original operator. Clearly the domain itself is a core, but it might be too big in certain applications like the present one.

We may assume that $H$ is formally given as above with some unknown dense domain $D(H)$, with $H_B$ and each $R_\alpha$ selfadjoint on certain dense domains $D(H_B)$ and $D(R_\alpha)$, which might be different, but with all  $H_B$ and $R_\alpha$ having a common core $\mathcal{C}$. This is the minimal assumption to make in order to have the sum definition of $H$ well-defined at all. Under this assumption the sum $\sum_{v\in V} vHv^\dagger$ is then also well-defined on $\mathbb{C}^d\otimes \mathcal{C}$ and its closure is exactly (an extension of) $\openone\otimes H_B$. For any $\psi\in \mathbb{C}^d\otimes \mathcal{C}$ the conditions of Chernoff's theorem, and in particular $(F(t)-\openone)(\psi)/t\rightarrow (\openone\otimes H_B) \psi$, are then satisfied, so (\ref{eq:converCher}) follows again.

Clearly if $H$ is self-adjoint with domain $\mathbb{C}^d\otimes D(H_B)$ then all $vHv^\dagger$ are also self-adjoint on that domain, but there are cases of $H$ with different domains, and that is when the above criterion with cores is needed.

We now discuss the question of how small $\Delta t$ needs to be to efficiently decouple. For bounded operators, \textit{the motion induced by the decoupling field needs to be faster than the fastest time-scale characterizing the unwanted interactions} \cite{LVioal1}. In the unbounded case, such a simple time-scale defined only by the interaction cannot be provided, as the convergence speed also crucially depends on the state, given by the speed of convergence of Chernoffs Theorem \eqref{eq:Chernoff}. Clearly there exist a $\tau(\psi,\epsilon)=\frac{t}{n}$ larger than zero for which $F(\tau)^{n}\psi$ is up to an error $\epsilon$ given by $\exp(t A)\psi$. Assuming that system and bath are initially uncorrelated, we may (through purification) without loss of generality assume that the initial bath state $\psi_B$ is pure. We can then define
 $\tau(\epsilon)=\text{inf}_{\psi_S} \tau(\psi_S \otimes \psi_B)>0$ as the critical time-scale for dynamical decoupling, where we used that the system space is finite-dimensional. This time-scale is harder to calculate than the finite-dimensional one, but we see a priori reasons why it should be much smaller than the latter.
\section{Conclusion}
So far we have considered the two extreme cases in which either extrinsic or intrinsic decoherence is present assuming the two mechanisms take place with the same decay rate. Clearly in an experimental situation both, a mixture $\mathcal{L} = \mathcal{L}_{\text{int}}+ \mathcal{L}_{\text{ext}}$ of extrinsic and intrinsic decoherence could be present. In this case, the asymptotic behavior of the gate error would be between those two extremal cases. It seems difficult to determine a general precise value, but estimates for the amount of intrinsic decoherence can be obtained based on the bounds 
$\|\overline{\mathcal{L}}_{\text{int}}\| \le \|\mathcal{L}_{\text{int}}\|$. The effective Lindbladian $\overline{\mathcal{L}}_{\text{int}}$ can be determined using process tomography. For intrinsic decoherence decay rates predicted by collapse models we are at present a few orders of magnitude away from the regime in which this becomes feasible. But with current advances in qubit design and a world-wide effort to increase the number of clean qubits this could come within reach soon.

Our results pave the way towards the experimental verification of alternative quantum theories (AQT) -- despite the presence of (extrinsic) decoherence. Even if the quantum noise is due to some unbounded coupling to an infinite dimensional environment we proved that the system evolution can be decoupled and hence distinguished from AQT. Furthermore, this decoupling of unbounded Hamiltonians has applications in quantum engineering beyond the scope of this paper. 
It is fascinating to contemplate that in the vast experimental evidence for dynamical decoupling such AQTs have already been discovered. The analysis of such experiments requires a detailed mathematical analysis, parts of which we have provided in \cite{USMath} and parts of it remain to be done in future.

\par

{\em Acknowledgements.} -- We thank Gavin Morley, Terry Rudolph, Y. Avron and Lorenza Viola for fruitful discussions. MF was supported by the Swiss National Science Foundation. DB acknowledges support from the EPSRC grant EP/M01634X/1

\bibliographystyle{apsrev}

\begin{thebibliography}{99}
\bibitem{Comment on domain initial state} Nevertheless, its time evolution is well defined due to the interaction being self-adjoint.
\end{thebibliography}

\end{document}